\documentclass[11pt,a4paper]{article}
\usepackage{amsmath,amssymb}
\usepackage{color}
\usepackage{graphicx}
\usepackage{subfig}
\usepackage{cite}                %%是引用格式变漂亮  %%PRD等APS模板必须禁用该项
\usepackage{hyperref}            %%超链接必须
\usepackage{multirow,makecell}
\usepackage{braket,bm}
\usepackage{cancel}
\RequirePackage{doi}
\usepackage{hyperref}
\usepackage{leftidx}
\usepackage[text={17cm,24.5cm},centering]{geometry}
\usepackage{textcomp}
\usepackage{wasysym}
\usepackage{mathtools}
\numberwithin{equation}{section}

\def \be {\begin{equation}}
\def \ee {\end{equation}}
\def \ba {\begin{array}}
\def \ea {\end{array}}
\def \bea{\begin{eqnarray}}
\def \eea{\end{eqnarray}}

\def \a {\alpha}

\def \g {\gamma}
\def \G {\Gamma}
\def \d {\delta}

\def \dg {\dagger}
\def \e {\epsilon}
\def \ve {\varepsilon}
\def \m {\mu}

\def \l {\lambda}
\def \L {\Lambda}

\def \s {\sigma}
\def \S {\Sigma}

\def \O {\Omega}

\def \t {\tau}
\def \z {\zeta}

\def \mD {\mathcal D}
\def \mE {\mathcal E}
\def \mF {\mathcal F}
\def \mG {\mathcal G}

\def \mI {\mathcal I}

\def \mR {\mathcal R}
\def \mS {\mathcal S}

\def \p {\partial}

\def \lt {\left}
\def \rt {\right}

\def \tr {\textrm{tr}}
\def \Tr {{\textrm{Tr}}}

\def \and {{\textrm{and}}}

\begin{document}
\begin{titlepage}
	
	\title{\textbf {R\'enyi negativities in non-equilibrium open free-boson chains}}
	\author{Hui-Huang Chen\footnote{chenhh@jxnu.edu.cn}~,}
	\date{}
	
	\maketitle
	\underline{}
	\vspace{-12mm}
	
	\begin{center}
		{\it
             College of Physics and Communication Electronics, Jiangxi Normal University,\\ Nanchang 330022, China\\
		}
		\vspace{10mm}
	\end{center}
	\begin{abstract}
	 In this paper, we consider the dynamics of R\'enyi negativities after a quantum quench in the free-boson chain with homogeneous dissipation. Initially we prepare the system in the squeezed thermal state, and then let it evolves under the tight-binding bosonic Hamiltonian with local linear dissipation. We use the Lindblad equation to solve the time evolution of the covariance matrix, from which one can obtain the time dependence of R\'enyi negativities. We are interested in the weak dissipation hydrodynamic limit where a quasi-particle picture emerges. In this limit, exact results of non-equilibrium dynamics of R\'enyi negativities can be obtained using the stationary phase method. We consider the R\'enyi negativities between both adjacent and disjoint regions in a infinite chain. We numerically test our analytical predictions and perfect matches have found.
	\end{abstract}
	
\end{titlepage}

\thispagestyle{empty}

\newpage

\tableofcontents
\section{Introduction}
In the past two decades, the exploration of different kinds of entanglement measures has made great progress in our understanding of quantum many-body systems, quantum field theory and quantum gravity. In condensed matter physics, entanglement is a powerful tool to characterize different phases of matter \cite{Amico:2007ag, Calabrese:2009qy, Eisert:2008ur}. In the AdS/CFT correspondence, the Ryu-Takayanagi formula \cite{Nishioka:2009un, Ryu:2006bv} firstly opens the route of understanding spacetime from entanglement and this idea turns out to have a key role in the black hole information loss paradox \cite{Hawking:1974sw, Hawking:1976ra, Almheiri:2020cfm}. In the studies of thermalizations of isolated quantum systems, people found that entanglement is a crucial quantity that characterizes how the subsystem reach equilibrium. Entanglement entropy or  Von Neumann entropy is the most useful entanglement measure to characterize the bipartite entanglement of a pure state. If we prepare our system in a pure state $\ket{\psi}$, the reduced density matrix (RDM) of the subsystem $A$ is obtained by tracing out degrees of freedom that are not in $A$, i.e. $\rho_A=\tr_{\bar A}\ket{\psi}\bra{\psi}$, where $\bar A$ is the complement of $A$. One can compute the von Neumann entropy from $\Tr\rho_A^{\a}$ via the replica trick \cite{Calabrese:2004eu}
\be
S_A\equiv-\Tr(\rho_A\log\rho_A)=\lim_{\a\rightarrow 1}S_A^{(\a)},
\ee
where $S_A^{(\a)}$ is the R\'enyi entropies
\be
S_A^{(\a)}=\frac{1}{1-\a}\log\Tr\rho_A^{\a}.
\ee
\par When one is interested in the entanglement between two subsystems $A_1$ and $A_2$ that are not complementary to each other, von Neumann entropy is no longer a good measure of entanglement. Since in this situation, $\rho_{A_1\cup A_2}$ is general a mixed state. Among different proposals, a computable measure of mixed state entanglement, logarithmic negativity turns out to be very useful \cite{Peres:1996dw, vidal2002computable, plenio2005logarithmic}. The definition is
\be
\mE=\log||\rho_A^{T_2}||,
\ee
where $||O||=\Tr\sqrt{O^{\dg}O}$ denotes the trace norm of the operator $O$ and $\rho_A^{T_2}$ is the partial transpose of RDM $\rho_A$ with respect to degree of freedom of subsystem $A_2$. Let $\ket{e_i^{(1)}}$ and $\ket{e_j^{(2)}}$ be two arbitrary bases of the Hilbert spaces associated to the degree of freedom on $A_1$ and $A_2$ respectively. The partial transpose (with respect to the second subsystem) of $\rho_A$ is defined as
\be
\bra{e_i^{(1)}e_j^{(2)}}\rho_A^{T_2}\ket{e_k^{(1)}e_l^{(2)}}=\bra{e_i^{(1)}e_l^{(2)}}\rho_A\ket{e_k^{(1)}e_j^{(2)}}.
\ee
If we denote the eigenvalues of $\rho_A^{T_2}$ as $\l_i$, we can write the trace norm of $\rho_A^{T_2}$ as
\be
||\rho_A^{T_2}||=\sum_{i}|\l_i|=\sum_{\l_i>0}\l_i-\sum_{\l_i<0}\l_i=1-2\sum_{\l_i<0}\l_i.
\ee
It's then useful to define the R\'enyi negativity
\be
\mathcal{\mE}_n=\log\Tr(\rho_A^{T_2})^n
\ee
which could be analytically continued from an even integer $n_e$ to obtain the logarithmic negativity, i.e. $\mE=\lim_{n_e\rightarrow 1}\mathcal{E}_{n_e}$. We must stress here that $\mE_n$ are not entanglement measures since they are not direct indicators of the sign of the eigenvalues of $\rho_A^{T_2}$.
\par The non-equilibrium dynamic of entanglement in the open quantum system has not been investigated much. In the paper \cite{alba2021spreading, carollo2022dissipative}, the authors first propose the quasi-particle picture in this context. In paper \cite{alba2022logarithmic, d2022logarithmic}, people explore the temporal evolution of fermionic logarithmic negativity in a free fermion chain. However, the quench dynamics of negativity in bosonic open quantum systems have not been studied before. In this manuscript, we will make some progress in this direction.
\par The quasi-particle picture of the entanglement after a global quantum quench has been proposed in \cite{Calabrese:2005in}. The key point is that one can assume that the initial state serves as the source of quasi-particles. Quasi-particle pairs emitted from the same point are highly entangled with opposite momenta $(k,-k)$. After the production, these quasi-particles travel ballistically with velocity $v_k=-v_{-k}$. The R\'enyi entropies of subsystem $A$ are proportional to the pairs of entangled quasi-particle shared with its complement at a given time $t$. For free models, we can write
\be\label{Snlat}
S_A^{(\a)}(t)=\int\frac{dk}{2\pi}s^{(\a)}_{\mathrm{GGE}}(k)\min(2|v_k|t,l),
\ee
with $l$ being the length of subsystem $A$ and $s^{(\a)}_{\mathrm{GGE}}(k)$ is momentum space density of the R\'enyi entropies in the Generalized Gibbs Ensemble (GGE) thermodynamic state \cite{vidmar2016generalized}.
\par The remaining part of this paper is organized as follows. In section 2, we introduce the problem we will study and review some basic facts about quantum systems with local linear dissipation. In section 3, we compute the time dependence of the covariance matrices with two different initial states. In section 4, we review the approach of computing R\'enyi entropies using the stationary phase approximation. In section 5, we apply this method to derive the exact formula of the time evolution of R\'enyi negativity in the weak dissipation hydrodynamic limit. We check our analytical predictions against exact numerical computation in section 6 and find perfect agreements. Finally, we conclude and prospect in section 7. The technical details of the squeezed thermal state are present in the appendix.
\section{Open quantum system}
\subsection{The tight-binding chain}
We consider a 1D lattice bosonic system with $L$ sites. At each site, one can introduce the creation $a_i^{\dg}$ and annihilation operators $a_i$, satisfying $[a_i,a_j^{\dg}]=\d_{ij}$. At the time $t=0$, we prepare the system in some initial state $\rho(0)$ and assume that at $t>0$ the system evolves according to the tight-binding bosonic chain
\be\label{H}
H=\epsilon\sum_{i=1}^La_i^{\dg}a_i-\frac{g}{2}\sum_{i=1}^{L}(a_i^{\dg}a_{i+1}+a_{i+1}^{\dg}a_i),
\ee
if the dissipation is absent.
The first term is the on-site energy of the particles, and the second terms describes the hopping of bosons between neighbouring sites.
In terms of canonical variables $x_n,p_n$ with $[x_i,p_j]=\mathrm{i}\d_{ij}$, the above Hamiltonian can be written as
\be
H=\frac12\sum_{i,j=1}^{2L}h_{ij}\xi_i\xi_j.
\ee
Here in $\xi$ we collect all $2L$ variables as $\xi=(x_1,p_1,\cdots,x_L,p_L)$ and $a_i=\frac{1}{\sqrt{2}}(x_i-\mathrm{i}p_i),a_i^{\dg}=\frac{1}{\sqrt{2}}(x_i+\mathrm{i}p_i)$.
From the Hamiltonian given in eq.~(\ref{H}), one find that $h_{ij}$ is a block-circulant matrix with elements given by
\be
h_{ij}=-\frac{g}{2}(\d_{i,j+1}+\d_{i,j-1})\begin{pmatrix}1&0\\0&1\end{pmatrix}+\e\d_{i,j}\begin{pmatrix}1&0\\0&1\end{pmatrix}.
\ee
For later convenience, we also introduce the $2L\times2L$ symplectic matrix $\Sigma$
\be\label{sigma}
[\xi_i,\xi_j]=\mathrm{i}\Sigma_{ij},\qquad \S_{ij}=\d_{ij}\sigma,\qquad \s=\begin{pmatrix}0&1\\-1&0\end{pmatrix}.
\ee
By translation invariance, we can make Fourier transform to (block) diagonalize the matrix $h_{ij}$ as
\be
\hat{h}_k\equiv\hat{h}_{kk}=\frac{1}{L}\sum_{i,j=1}^{L}e^{\mathrm{i}k(i-j)}h_{ij}=\left(\e-g\cos k\right)\mathbb{I}_2.
\ee
where $\mathbb{I}_2$ is the $2\times 2$ identity matrix.
\par The tight-binding Hamiltonian eq.~(\ref{H}) can be diagonalized by Bogoliubov transform
\be
b_k=\frac{1}{\sqrt{L}}\sum_{i=1}^{L}e^{\mathrm{i}kn}a_i.
\ee
In terms of these new Bogoliubov modes $b_k$, the Hamiltonian eq.~(\ref{H}) becomes
\be
H=\sum_{k}\varepsilon(k)b_k^{\dg}b_k.
\ee
with dispersion and quasi-particle velocity given by
\be
\varepsilon(k)=\epsilon-g\cos k,\qquad v(k)=\frac{\p\varepsilon(k)}{\p k}=g\sin k.
\ee
\subsection{The Lindblad dynamics}
In this section and the following part, we will study the non-equilibrium dynamics of entanglement in the tight-binding bosonic chain eq.~(\ref{H}) with homogeneous gain and loss dissipation. For this open quantum system, we adopt the Lindblad master equation approach \cite{lindblad1976generators, breuer2002theory}. The time evolution of the density matrix is governed by the Lindblad equation
\be\label{Lindblad}
\frac{d}{dt}\rho(t)=-\mathrm{i}[H,\rho]+\sum_{j=1}^L\sum_{\a=\pm}L_j^{(\a)}\rho L_j^{(\a)\dg}-\frac12\{L_j^{(\a)\dg}L_j^{(\a)},\rho\}.
\ee
In this paper, we are interested in the gain and loss processes, i.e. $L_j^{(+)}=\sqrt{\g^+}a_j^{\dg}$ and $L_j^{(-)}=\sqrt{\g^-}a_j$. In terms of canonical variables $\xi_j$, the above equation is equivalent to
\be
\frac{d}{dt}\rho(t)=-\mathrm{i}[H,\rho]+\sum_{i,j=1}^{2L}C_{ij}\big(\xi_i\rho\xi_j-\frac12\{\xi_i\xi_j,\rho\}\big),
\ee
where the elements of the $2L\times2L$ matrix $C$ are given by
\be\label{C}
C_{ij}=C_{ij}^{(+)}+C_{ij}^{(-)},\qquad C_{ij}^{(\pm)}=\frac12\d_{ij}\g^{\pm}\begin{pmatrix}1&\mp\mathrm{i}\\\pm\mathrm{i}&1\end{pmatrix}.
\ee
In general, $C_{ij}$ can be long-ranged. Here we only consider the local dissipation, i.e. each site interacts with the environment independently.
Since the model is free and undergoes a Gaussian Markovian open quantum time-evolution, which means Gaussian states remain Gaussian when the system evolves in time. Thus the properties of the system are fully characterized by the covariance matrix
\be
G_{ij}(t)\equiv\frac12\Tr[\rho(t)\{\xi_i,\xi_j\}].
\ee
Using the Lindblad equation (\ref{Lindblad}), one can solve $G(t)$. In terms of its symbol $\hat{G}_k(t)$, we write
\be
\hat{G}_k(t)=\hat{G}_k^{(1)}(t)+G_k^{(2)}(t)
\ee
where \cite{carollo2022dissipative}
\be\label{hatGk}
\hat{G}_k^{(1)}(t)=e^{t\L_k}\hat{G}_k(0)e^{t\L^{\dg}_k},\qquad \hat{G}_k^{(2)}(t)=\bar\g\int_0^tdt'e^{t'\L_k}e^{t'\L^{\dg}_k},
\ee
with
\be
\L_k=\s\hat{h}_k-\frac12\g\mathbb{I}_2.
\ee
Here we have defined
\be
\bar\g=\frac{\g^++\g^-}{2},\qquad \g=\g^--\g^+.
\ee
It's clear that in our case, $\bar\g$ and $\g$ do not dependent on $k$, which turns out to be only valid in local dissipations. Since the entanglement entropy is related to the symplectic eigenvalues of $G$, then it's useful to introduce the matrix $\G$ defined as $\G\equiv\mathrm{i}\S G$.
\par The term $\hat{\G}_k^{(2)}(t)$ is universal which means it doesn't depend on the initial state, while $\G_k^{(1)}(t)$ contains information on initial state. In our case, since $[\L_k,\L_k^{\dg}]=0$, $\hat{G}_k^{(2)}(t)$ hence $\hat{\G}_k^{(2)}(t)$ can be easily computed. As a result, $\hat{\G}_k^{(2)}(t)$ takes the very simple form
\be\label{Ga2}
\hat{\G}_k^{(2)}(t)=-\frac{\bar\g}{\g}(1-e^{-\g t})\s_y.
\ee

\section{Quench dynamics}
In this part, we will consider a one-dimensional bosonic system at $t<0$ prepared in some initial state, and the system evolves under the tight-binding Hamiltonian with gain and loss dissipation at $t>0$. We will first study the simple case where the initial state is a thermal state. In this particular setting, the computation is straightforward and easy, but we will introduce some important concepts and quantities which are crucial for subsequent studies. Then we will focus on the case where our system was initially prepared in a squeezed thermal state \cite{carollo2022dissipative}. A squeezed thermal state has more interesting properties. Due to its complications, we will need a more sophisticated method to obtain the dynamics of entanglement. We are interested in the weak-dissipation hydrodynamic limit: $t,l\rightarrow\infty$,$\g^{\pm}\rightarrow 0$ with $t/l$ and $\g^{\pm}t$ keep fixed \cite{carollo2022dissipative, alba2022logarithmic, alba2022hydrodynamics}.
\subsection{Thermal states}
Let first consider a simple example where the initial state is a thermal state
\be
\rho_0=\rho_{th}.
\ee
It's convenient to introduce the Fock space basis $\ket{\mathbf{n}}\equiv\otimes_{i}\ket{n_i}$, defined by products of eigenstates of the number operator of each site. In this basis, the thermal state $\rho_{th}\propto e^{-\beta H(g=0)}$ can be written as
\be
\rho_{th}=\sum_{\mathbf{n}}\prod_{i}(1+\bar n)^{-1}\left(\frac{\bar n}{1+\bar n}\right)^{n_i}\ket{\mathbf{n}}\bra{\mathbf{n}},
\ee
with $\bar n=(e^{\beta\e}-1)^{-1}$ is the average occupation number of per-site at the temperature $T=1/\beta$.
The initial covariance matrix is
\be
G(0)_{i,j}=\eta\d_{ij}\begin{pmatrix}1&0\\0&1\end{pmatrix},\quad \text{with}\quad \eta\equiv \bar n+\frac12
\ee
from which one can easily obtain
\be
\hat{G}_k^{(1)}=\eta e^{-\g t}\mathbb{I}_2
\ee
and
\be
\hat{G}_k^{(2)}=\frac{\bar\g}{\g}(1-e^{-\g t})\mathbb{I}_2.
\ee
Then we can write $\hat{\G}_k$ as
\be
\hat{\G}_k(t)=-(n_k(t)+\frac12)\s_y
\ee
where we have introduced the particle density in momentum space: $n_k(t)\equiv\tr(\rho(t)b_k^{\dg}b_k)$.
Then it's clear that $n_k$ is dependent on time in contrast with the case when there is no dissipation. When the dissipation is absent, each mode densities $n_k$ are conserved quantities which are constitutions of the GGE.  In our case, one can find out the time dependence of $n_k$ from the Lindblad equation. It turns out that $n_k(t)$ satisfies the rate equation \cite{carollo2022dissipative}
\be\label{rate}
\frac{d}{dt}n_k(t)=-\g n_k+\g^+,
\ee
whose solution is given by
\be
n_k(t)=e^{-\g t}n_k(0)+\frac{\g^+}{\g}(1-e^{-\g t})
\ee
with
\be
n_k(0)=\bar n.
\ee
In this case, since $n_k(t)$ is independent on $k$, $\hat{\G}_k(t)$ also does not dependent on $k$. One can easily find that $\G(t)$ is block-diagonal with elements given by
\be
\G_{ij}(t)=\frac{1}{L}\sum_{k}e^{\mathrm{i}(i-j)k}\hat{\G}_k(t)=-\d_{ij}(n_k(t)+\frac12)\s_y.
\ee
Obviously, the spectrum of $\G(t)$ is $\{\pm(n_k(t)+\frac12)\}$. If we denote $2l\times 2l$ matrix $\G_A$ as the full matrix $\G_{ij}$ with the index $i,j$ restricted to subsystem $A$. Then it's clear that the trace of odd powers of $\G_A$ vanish. Only even powers of $\G_A$ have non-zero trace.
\par In this case, $\hat{\G}_k(t)$ is actually independent of $k$, we can easily find the analytic result of $\Tr[(\G_A)^{2n}]$. In fact, for arbitrary function $\mF(z)$ analytic around $z=0$, we have
\be
\Tr\mF(\G^2_A)=l\int_0^{2\pi}\frac{dk}{2\pi}\Tr\mF(\hat{\G}_k^2)=l~\Tr\mF(\hat{\G}_k^2).
\ee
If we want to compute the time evolution of the R\'enyi entropies, we should choose
\be
\mF_{\a}(z)=-\frac12\frac{1}{1-\a}\log\left[\left(\sqrt{z}+\frac12\right)^{\a}-\left(\sqrt{z}-\frac12\right)^{\a}\right].
\ee
To obtain the von Neumann entropy, one should instead choose
\be
\mF(z)=-\frac12\left[\left(\sqrt{z}+\frac12\right)\log\left(\sqrt{z}+\frac12\right)-\left(\sqrt{z}-\frac12\right)\log\left(\sqrt{z}-\frac12\right)\right].
\ee
Thus the R\'enyi entropies with the index $\a$ is
\be\label{logZP}
\begin{split}
&S_A^{(\a)}(t)=2l\mF_{\a}((n_k(t)+1/2)^2)\\
&=\frac{l}{1-\a}\log[(n_k(t)+1)^{\a}-n_k(t)^{\a}].
\end{split}
\ee
The time evolution of the von Neumann entropy is simply given by
\be\label{SA}
S_A(t)/l=(n_k(t)+1)\log(1+n_k(t))-n_k(t)\log n_k(t).
\ee
\subsection{Squeezed thermal state}
In this section, we will consider the case where the initial state is a squeezed thermal state defined by \cite{calzetta2009nonequilibrium}
\be
\rho(0)=\mS(r,\phi)\rho_{th}\mS^{\dg}(r,\phi),
\ee
where $\mS(r,\phi)$ is the squeezing operator defined by
\be\label{squ}
\mS(r,\phi)=\prod_{i}\exp[\frac{r}{2}(a_i^2e^{-2\mathrm{i}\phi}-a_i^{\dg 2}e^{2\mathrm{i}\phi})].
\ee

\par For simplicity, we consider the case in which the squeeze parameters are uniform on each site $r_i=r\in\mathbb{R}$ and $\phi_i=0$.
Using the property of the squeezed thermal state (see the appendix), one could find that the initial covariance matrix is block-diagonal in position space with elements given by
\be
G_{ij}(0)=\eta\begin{pmatrix}e^{2r}&0\\0&e^{-2r}\end{pmatrix}\d_{ij},
\ee
where $\eta\equiv\bar n+1/2$, see the appendix for details. The symbol of $G(0)$ is given by
\be
\hat{G}_k(0)=\eta\begin{pmatrix}e^{2r}&0\\0&e^{-2r}\end{pmatrix}.
\ee
\par Then from eq.~(\ref{hatGk}), it's straightforward to obtain $\hat{\G}_k^{(1)}(t)$ as
\be
\hat{\G}_k^{(1)}(t)=-\eta e^{-\g t}[\mathrm{i}\cos(2\ve_kt)\sinh(2r)\s_x+\cosh(2r)\s_y+\mathrm{i}\sin(2\ve_k t)\sinh(2r)\s_z].
\ee
For later's convenience, we rewrite it as
\be
\hat{\G}_k^{(1)}(t)=-\eta e^{-\g t}[\cosh(2r)\s_y+\mathrm{i}\sinh(2r)\s_x e^{-2\mathrm{i}\ve_kt\s_y}].
\ee
Including the contribution of $\hat{\G}_k^{(2)}(t)$ in eq.~(\ref{Ga2}), $\hat{\G}_k(t)$ has the form
\be
\hat{\G}_k(t)=a\s_y+b\s_xe^{-2\mathrm{i}\ve_kt\s_y},
\ee
where
\be\label{a}
a=-\frac{\bar\g}{\g}\left(1-e^{-\g t}\right)-\eta\cosh(2r)e^{-\g t},
\ee
and
\be\label{B}
b=-\mathrm{i}\eta\sinh(2r)e^{-\g t}.
\ee
\par In this case, $\hat{\G}_k(t)$ does depend on $k$ and we need more sophisticated approaches to evaluate $\Tr[(\G_A)^{2n}]$ and $\Tr[(\G_A^{T_2})^{2n}]$.
\section{Entanglement entropy}
In this section, as a warm-up, let's quickly review the method of computing the evolution of entanglement entropy in the free boson chain with linear dissipation. Since this method was widely used in the literature, see for example \cite{calabrese2012quantum, alba2022logarithmic}, here we just review the main point of the strategy and in the next section, we will apply this method to compute the dynamics of R\'enyi negativities in our tight-binding bosonic chain with dissipation. Subsystem $A$ consists of $l$ continuous sites and $\bar A$ is complementary. In this part, we follow section 3 of the paper \cite{calabrese2012quantum} closely.
\par The matrix $\G_A$ are obtained from $\G$ with column and row restricted in $A$
\be
\G_{mn}=\int\frac{dk}{2\pi}e^{\mathrm{i}(m-n)k}\hat{\G}_k,\qquad m,n=1,2,\cdots,l.
\ee
Here we assume that $A=[1,l]$. From the equation above, it's straightforward to get
\be
\Tr[(\G_A)^{2n}]=\int\prod_{j=0}^{2n-1}\frac{dk_j}{2\pi}\sum_{m=1}^{l}e^{-\mathrm{i}m(k_j-k_{j+1})}
~\Tr\prod_{i=0}^{2n-1}\hat{\G}_{k_i}.
\ee
Using the following formula
\be\label{sum}
\sum_{m=1}^{l}e^{-\mathrm{i}mk}=\frac{l}{2}\int_{-1}^{+1}du w(k)e^{\mathrm{i}(lu-l-1)k/2},\qquad w(k)=\frac{k}{2\sin(k/2)},
\ee
we can rewrite $\Tr[(\G_A)^{2n}]$ as
\be
\Tr[(\G_A)^{2n}]=\left(\frac{l}{2}\right)^{2n}\int\displaylimits_{[0,2\pi]^{2n}}\frac{d^{2n}k}{(2\pi)^{2n}}\int\displaylimits_{[-1,1]^{2n}}d^{2n}u D(\{k\})F(\{k\})e^{\mathrm{i}l\sum_{j=0}^{2n-1}u_j(k_{j}-k_{j+1})/2},
\ee
where
\be
D(\{k\})=\prod_{j=0}^{2n-1}\frac{k_{j+1}-k_j}{2\sin[(k_{j+1}-k_j)/2]},
\ee
and
\be\label{F}
F(\{k\})=\Tr\prod_{i=0}^{2n-1}\hat{\G}_{k_i}.
\ee
\par The trace over a product of $\hat{\G}_{k_i}$ has already been computed in \cite{calabrese2012quantum}. Firstly, the product is
\be
\prod_{i=0}^{2n-1}\hat{\G}_{k_i}=\sum_{p=0}^{2n}(-1)^pa^{2n-p}b^p(\mathrm{i}\s_z)^p\sum_{0\leq j_1<j_2<\cdots<j_p\leq 2n-1}(-1)^{\sum_{i=1}^pj_i}e^{-2\mathrm{i}t\sum_{i=1}^p(-1)^{p-i}\varepsilon(k_{j_i})\s_y}.
\ee
Taking the trace, one get
\be\label{F1}
\Tr\prod_{i=0}^{2n-1}\hat{\G}_{k_i}=\sum_{p=0}^{n}(-1)^pa^{2n-2p}b^{2p}\sum_{0\leq j_1<j_2<\cdots<j_{2p}\leq 2n-1}(-1)^{\sum_{i=1}^{2p}j_i}2\cos\Big[2t\sum_{i=1}^{2p}(-1)^{i}\varepsilon(k_{j_i})\Big].
\ee
Using the invariant of the integral under the permutation of the quasimomenta, we can write
\be
\Tr\prod_{i=0}^{2n-1}\hat{\G}_{k_i}=\sum_{p=0}^{n}\binom{n}{p}a^{2n-2p}b^{2p}2\cos\Big[2t\sum_{i=0}^{2p-1}(-1)^{i}\varepsilon(k_{i})\Big].
\ee
Because of the measure $\m(\{\z_j\})$ is symmetric with respect to $\vec{\z}\rightarrow -\vec{\z}$, we can replace the cosine in the equation above with a phase
\be
F(\{k\})\rightarrow 2\sum_{p=0}^{n}\binom{n}{p}a^{2n-2p}b^{2p}e^{2\mathrm{i}t\sum_{i=0}^{2p-1}(-1)^{i}\varepsilon(k_{i})}.
\ee
\par It's convenient to define
\be\label{zeta}
\begin{split}
&\zeta_0=u_0\\
&\zeta_j=u_{j}-u_{j-1},\qquad j\in[1,2n-1].
\end{split}
\ee
Then
\be
\Tr[(\G_A)^{2n}]=\left(\frac{l}{2}\right)^{2n}\int\displaylimits_{[0,2\pi]^{2n}}\frac{d^{2n}k}{(2\pi)^{2n}}\int\displaylimits_{R_{u}}d^{2n}\zeta D(\{k\})F(\{k\})e^{\mathrm{i}l\sum_{j=1}^{2n-1}\zeta_j(k_{j}-k_{0})/2},
\ee
where of domain of the integral $R_{u}$ is defined by
\be
-1\leq \sum_{j=0}^p\zeta_j\leq 1,\qquad \forall p\in[0,2n-1].
\ee
Since we are considering the case $l\gg 1$, we can use stationary point approximation to evaluating the above integral. The stationarity with respect to the variables $\zeta_0,\zeta_1,\cdots,\zeta_{2n-1}$ implies that
\be
k_j=k_0,\qquad \forall j\in[1,2n-1].
\ee
\par We can replace any $k_j$ with $k_0$ except for highly oscillating terms: $e^{-2\mathrm{i}\varepsilon_{k_i}t\s_y}$ in $\hat{\G}_{k_i}$. By this rule, one can simply dropping the term $D(\{k\})$ in the integral since $D(\{k\})=1$ at the stationary point.
Moreover, the above integrand does not depend on $\zeta_0$, we can first integrate out $\zeta_0$ obtaining
\be
\Tr[(\G_A)^{2n}]=\left(\frac{l}{2}\right)^{2n}\int\displaylimits_{[0,2\pi]^{2n}}\frac{d^{2n}k}{(2\pi)^{2n}}\int d^{2n-1}\zeta ~\mu(\{\zeta\})F(\{k\})e^{\mathrm{i}l\sum_{j=1}^{2n-1}\zeta_j(k_{j}-k_{0})/2}.
\ee
Here the function $\m(\{\zeta\})$ is the measure of the integral of the variables $\z_1,\z_2,\cdots,\z_{2n-1}$
\be\label{mu}
\m(\{\zeta\})=\max\Big[0,\min_{j\in[0,2n-1]}\Big[1-\sum_{i=0}^j\z_i\Big]+\min_{j\in[0,2n-1]}\Big[1+\sum_{i=0}^j\z_i\Big]\Big].
\ee
This measure is obviously symmetric under the change of variables $\vec{\z}\rightarrow-\vec{\z}$.
\par To apply stationary phase approximation, we write
\be\label{TrGaint}
\Tr[(\G_A)^{2n}]=l\left(\frac{l}{2}\right)^{2n-1}\sum_{p=0}^{n}\binom{n}{p}\int\frac{dk_0}{2\pi}a^{2n-2p}b^{2p}\L_{n;p}(k_0),
\ee
where
\be
\L_{n;p}(k_0)=\int\displaylimits_{[0,2\pi]^{2n}}\frac{d^{2n}k}{(2\pi)^{2n}}\int d^{2n-1}\zeta ~\mu(\{\zeta\})e^{\mathrm{i}l\sum_{j=1}^{2n-1}\zeta_j(k_{j}-k_{0})/2+2\mathrm{i}t\sum_{j=0}^{2p-1}(-1)^j\varepsilon_j}.
\ee
The stationary point is
\be
\begin{split}
&k_j^*=k_0,\qquad\qquad \qquad j=1,2,\cdots,2n-1\\
&\z_j^*=-(-1)^j4\varepsilon'(k_0) t/l,\qquad j=1,2,\cdots,2p-1\\
&\z_j^*=0 \qquad\qquad \qquad j=2p,\cdots,2n-1.
\end{split}
\ee
Using the definition of the measure given in eq.~(\ref{mu}), one could obtain the measure at the stationary point $\m(\{\z_j^*\})$=2 when $p=0$ and $\m(\{\z_j^*\})=2g_1(k_0,t/l)$ for $p\neq 0$, where
\be
g_1(k,x)=\max(1-2|v(k)|x,0).
\ee
\par We now use the formula of multi-dimensional stationary phase approximation for large $l$
\be
\int_{\mD}d^Nxp(\mathbf{x})e^{\mathrm{i}lq(\mathbf{x})}=\left(\frac{2\pi}{l}\right)^{N/2}p(\mathbf{x}_0)|\det M|^{-1/2}\exp\Big[\mathrm{i}lq(\mathbf{x}_0)+\frac{\mathrm{i}\pi\s_M}{4}\Big].
\ee
Here $\mathbf{x}_0$ is the stationary point determined by the condition $\nabla q(\mathbf{x}_0)=0$, $M_{ij}=\p_{x_i}\p_{x_j}q(\mathbf{x}_0)$ is the Hessian matrix of the function $q(\mathbf{x})$ evaluated at $\mathbf{x}_0$. $\s_M$ is the signature of the matrix $M$ which is zero in this case and $\det M=-4^{1-2n}$.
\par Therefore
\be
\L_{n;p}(k_0)=\left(\frac{2}{l}\right)^{2n-1}\begin{cases}2g_1(k_0,t/l),\quad p\neq 0\\
2\quad\qquad\qquad\quad p=0\end{cases}
\ee
Substituting the above equation to eq.~(\ref{TrGaint}) and do the summation, we finally obtain
\be\label{TrGaf}
\Tr[(\G_A)^{2n}]=2la^{2n}+2l[(a^2+b^2)^n-a^{2n}]\int_0^{2\pi}\frac{dk}{2\pi}\max(0,1-2|v(k)|t/l).
\ee
For R\'enyi entropy with index $\a$, we have
\be
S^{(\a)}_A=2l\mF_{\a}(a^2)+2l\big[\mF_{\a}(a^2+b^2)-\mF_{\a}(a^2)\big]\int_0^{2\pi}\frac{dk}{2\pi}\max(0,1-2|v(k)|t/l).
\ee
The expression of entanglement entropy has the same form with $\mF_{\a}$ replaced by $\mF$.
\subsection{Interpretation}
As discussed in the previous subsection, one conclude that the multi-dimensional stationary phase method can be used to derive the time evolution of R\'enyi entropies in the weak-dissipation hydrodynamic limit. From eq.~(\ref{TrGaf}), we have \cite{carollo2022dissipative}
\be\label{TrF}
\Tr(\mathcal{F}(\G_A^2))=\int_{0}^{2\pi}\frac{dk}{2\pi}[2\mF(a^2)-\Tr\mF(\hat{\G}_k^2)]\min(l,2|v_k|t)+l\int_0^{2\pi}\frac{dk}{2\pi}\Tr\mF(\hat{\G}_k^2).
\ee
\par We define the particle density in momentum space as
\be
n_k=-a-\frac12.
\ee
If we substitute the expression of $a$ (cf. eq.~(\ref{a})), then it's easy to check that $n_k(t)$ indeed satisfies the rate equation (\ref{rate}).
Thus we also have
\be
n_k(t)=e^{-\g t}n_k(0)+\frac{\g^+}{\g}(1-e^{-\g t}),
\ee
where the initial density is given by eq.~(\ref{n0})(see the appendix for details)
\be
n_k(0)=\bar n\cosh(2r)+\sinh^2r.
\ee
From eq.~(\ref{TrF}), we can write the R\'enyi entropies as \cite{carollo2022dissipative}
\be\label{Salpha}
S_A^{(\a)}(t)=\int\frac{dk}{2\pi}[s^{(\a)}_{\mathrm{q}}(k)-s^{(\a)}_{\mathrm{mix}}(k)]\min(l,2|v(k)|t)+l\int\frac{dk}{2\pi}s^{(\a)}_{\mathrm{mix}}(k),
\ee
where $s^{(\a)}_{\mathrm{q}}(k)$ is defined as
\be\label{sq}
s^{(\a)}_{\mathrm{q}}(k)=2\mF_{\a}(a^2)=-\log[(1+n_k)^{\a}-n_k^{\a}],
\ee
and
\be\label{sm}
\begin{split}
s^{(\a)}_{\mathrm{mix}}(k)=\Tr\mF_{\a}(\hat{\G}_k^2)=2\mF_{\a}(a^2+b^2).
\end{split}
\ee
\par The first term in eq.~(\ref{Salpha}) describes the correlations between each pair of quasi-particles. Comparing to the case without dissipation, here an additional term $-s^{(\a)}_{\mathrm{mix}}(k)$ in the square bracket appears which indicates the evolution is non-unitary. In the limit $\g t\rightarrow\infty$, we have $b\rightarrow 0$, $n_k\rightarrow n_{\infty}\equiv\g^+/\g$, and $s^{(\a)}_{\mathrm{mix}}\rightarrow s^{(\a)}_{\mathrm{q}}$, thus the first term vanishes, and only the last term survives which indicates the incoherent action of the environment. Due to the contribution from the last term in eq.~(\ref{Salpha}), at $t=0$, we have non-zero values for the R\'enyi entropies. Indeed, at $t=0$, since the first term in eq.~(\ref{Salpha}) vanishes, one has that $S_A^{(\a)}(t=0)=2l\mF_{\a}(\eta^2)=-l/(1-\a)\log[(1+\bar n)^{\a}-\bar n^{\a}]$.
\par Note that in the gain and loss dissipation, $\g$ actually does not depend on $k$, so as to $a$ and $b$. Actually the integration of $k$ in eq.~(\ref{Salpha}) can be done analytically in this case. Introducing the function
\be\label{mG}
\begin{split}
&\mG(x)=\int_0^{2\pi}\frac{dk}{2\pi}\min(1,2|\sin(k)|x)\\
&=\Theta(1/2-x)\frac{4x}{\pi}+\Theta(x-1/2)\frac{2}{\pi}(2x-\sqrt{4x^2-1}+~\mathrm{arcsec}(2x)),
\end{split}
\ee
where $\Theta(x)$ is the Heaviside step function, we can obtain exact results for the time evolution of R\'enyi entropies as
\be
S_A^{(\a)}(t)/l=2\mF_{\a}(a^2)\mG(|g|t/l)+2\mF_{\a}(a^2+b^2)[1-\mG(|g|t/l)].
\ee
\section{R\'enyi negativities}
In this section, we consider the case where the subsystem $A$ is made of two disjoint regions, i.e. $A=A_1\cup A_2$. Here for simplicity, we assume that both the length of $A_1$ and $A_2$ are $l_0$. Obviously, the length of the subsystem $A$ is $l\equiv 2l_0$. We will adopt the method sketched in the last section to calculate the time evolution of entanglement negativity of subsystem $A$ in the weak dissipation hydrodynamic limit.
\par In this case, considering the block structure of the covariance matrix, we can write it as
\be
G_A=\begin{pmatrix}G^{11}&G^{12}\\G^{21}&G^{22}\end{pmatrix},
\ee
where 1 and 2 are the labels of the sites in $A_1$ and $A_2$ respectively.
\par To compute the entanglement negativity, one involves considering the partially transposed RDM with respect to the second region of the subsystem. The effect of the partial transposing with respect to the second part of the subsystem can be encoded in the covariance matrix. If we denote the corresponding covariance matrix as $G_A^{T_2}$, then we have
\be\label{GA}
G_A^{T_2}=\begin{pmatrix}I_1&O\\O&R_2\end{pmatrix}\begin{pmatrix}G^{11}&G^{12}\\G^{21}&G^{22}\end{pmatrix}\begin{pmatrix}I_1&O\\O&R_2\end{pmatrix}
=\begin{pmatrix}G^{11}&G^{12}R_2\\R_2G^{21}&R_2G^{22}R_2\end{pmatrix},
\ee
where $I_1=\mathbb{I}_{l_0}$ is the $l_0\times l_0$ identity matrix in the first region, and the effect of $R_2$ is to reflect the momentum in the second region, i.e. $(R_2)_{ij}=\d_{ij}\s_z$.
\par Similarly, we can write the matrix $\G_A=\mathrm{i}\S_A G_A$ in a block form
\be\label{GaA}
\G_A=\begin{pmatrix}\G^{11}&\G^{12}\\\G^{21}&\G^{22}\end{pmatrix}.
\ee
\begin{figure}
        \centering
        \subfloat
        {\includegraphics[width=7cm]{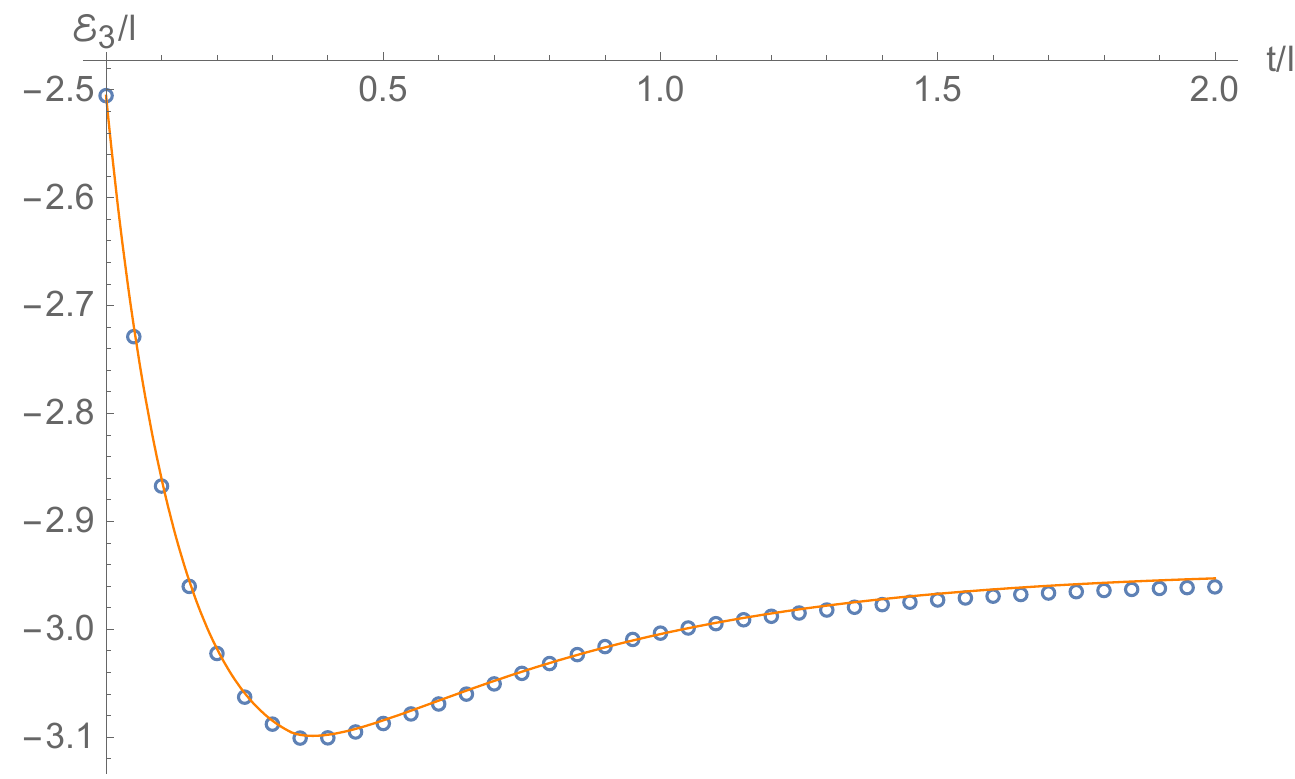}} \quad\quad
        {\includegraphics[width=7cm]{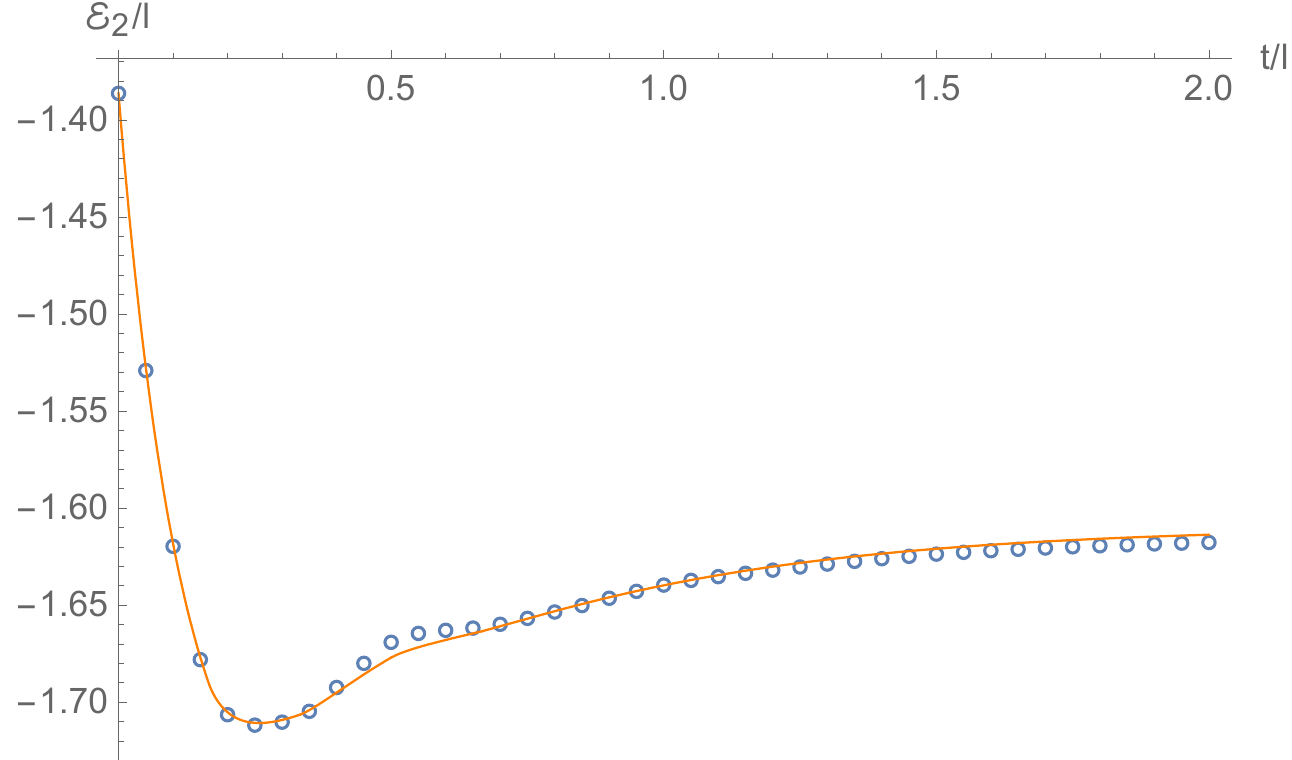}}
        \caption{R\'enyi negativity $\mathcal{E}_{\a}$ as a function of $t/l$. The full lines are the quasi-particle predictions (cf. eq.~(\ref{Ea}) and eq.~(\ref{Ed})).Left panel: $\mathcal{E}_{3}$ for adjacent intervals. Right panel: $\mathcal{E}_{2}$ for disjoint intervals with $d=100$. The parameters we choose are $L=300,l_0=50$, $\g^+=2l_0^{-1},\g^-=3l_0^{-1},\e=0.2,g=1.5,r=0.5,\eta=2$. As shown in the figure, the agreement is perfect.}
        \label{fig1}
\end{figure}
\subsection{Adjacent interval}
We first consider the case where $A_1$ and $A_2$ are adjacent. Then the formula for the disjoint case can be obtained similarly.
\par In the adjacent case, the matrix $\G_A$ is related to the symbol $\hat{\G}_k$ as
\be
(\G_A)_{mn}=\int\frac{dk}{2\pi}e^{\mathrm{i}(m-n)k}\hat{\G}_k,\qquad m,n=1,2,\cdots,l
\ee
Then each block in eq.~(\ref{GaA}) are $l_0\times l_0$ matrices whose elements are again $2\times 2$ matrices. We have
\be
\begin{split}
&\G^{11}_{mn}=\G^{22}_{mn}=\int\frac{dk}{2\pi}e^{\mathrm{i}(m-n)k}\hat{\G}_k,\qquad m,n=1,2,\cdots,l_0\\
&\G^{12}_{mn}=\int\frac{dk}{2\pi}e^{-\mathrm{i}l_0k}e^{\mathrm{i}(m-n)k}\hat{\G}_k,\qquad m,n=1,2,\cdots,l_0\\
&\G^{21}_{mn}=\int\frac{dk}{2\pi}e^{\mathrm{i}l_0k}e^{\mathrm{i}(m-n)k}\hat{\G}_k,\qquad m,n=1,2,\cdots,l_0
\end{split}
\ee
where we have rearranged the range of the column and row of each block to $1,2,\cdots,l_0$, and some phase must be included in $\G^{12}$ and $\G^{21}$. It turns out that these phases are crucial in determining the stationary points.
\par If we denote the $\G$ matrix associated the matrix $G^{T_2}$ as $\G^{T_2}$, then we have
\be
\G_A^{T_2}=\mathrm{i}\Sigma_A G_A^{T_2}=\begin{pmatrix}\G^{11}&\G^{12}R_2\\-R_2\G^{21}&-R_2\G^{22}R_2\end{pmatrix}
\equiv\begin{pmatrix}\tilde{\G}^{11}&\tilde{\G}^{12}\\\tilde{\G}^{21}&\tilde{\G}^{22}\end{pmatrix}.
\ee
Here the minus signs reflect the fact $\{\s,\s_z\}=0$ recalling that $\s=-\mathrm{i}\s_y$ defined in eq.~(\ref{sigma}).
\par Then the following relations must be hold
\be
\begin{split}
&\tilde{\G}^{11}_{mn}=\int\frac{dk}{2\pi}e^{\mathrm{i}(m-n)k}\hat{\G}_k,\qquad m,n=1,2,\cdots,l_0\\
&\tilde{\G}^{12}_{mn}=\int\frac{dk}{2\pi}e^{-\mathrm{i}l_0k}e^{\mathrm{i}(m-n)k}\hat{\G}_k\s_z,\qquad m,n=1,2,\cdots,l_0\\
&\tilde{\G}^{21}_{mn}=-\int\frac{dk}{2\pi}e^{\mathrm{i}l_0k}e^{\mathrm{i}(m-n)k}\s_z\hat{\G}_k,\qquad m,n=1,2,\cdots,l_0\\
&\tilde{\G}^{22}_{mn}=-\int\frac{dk}{2\pi}e^{\mathrm{i}(m-n)k}\s_z\hat{\G}_k\s_z,\qquad m,n=1,2,\cdots,l_0
\end{split}
\ee
It's easy to see that the trace of odd powers of $\G_A^{T_2}$ vanish. We compute the trace of first three even powers of $\G_A^{T_2}$ by brute force, and conjectured the following formula is valid for general integer $n$
\be
\Tr[(\G_A^{T_2})^{2n}]=\int\prod_{i=0}^{2n-1}\frac{dk_i}{2\pi}(-1)^n\prod_{j=0}^{2n-1}2\sin\left[\frac{(k_j-k_{j+1})l_0}{2}\right]\sum_{m=1}^{l_0}e^{-\mathrm{i}m(k_j-k_{j+1})}
~\Tr\prod_{i=0}^{2n-1}\hat{\G}_{k_i}.
\ee
In principle, the formula above can be proved for example by induction. But we leave the details of the proof for the interested readers. Then following the same strategy in computing the entanglement entropy in the last section, we use the formula eq.~(\ref{sum}) to rewrite $\Tr[(\G_A^{T_2})^{2n}]$ as
\be
\Tr[(\G_A^{T_2})^{2n}]=\left(\frac{l_0}{2}\right)^{2n}\int\displaylimits_{[0,2\pi]^{2n}}\frac{d^{2n}k}{(2\pi)^{2n}}\int\displaylimits_{[-1,1]^{2n}}d^{2n}u D(\{k\})\tilde{F}(\{k\})e^{\mathrm{i}l_0\sum_{j=0}^{2n-1}u_j(k_{j}-k_{j+1})/2},
\ee
where in this case $\tilde{F}(\{k\})$ is a product of two complicated factors
\be\label{tF}
\tilde{F}(\{k\})=(-1)^n\prod_{j=0}^{2n-1}2\sin\left[\frac{(k_j-k_{j+1})l_0}{2}\right]\times\Tr\prod_{i=0}^{2n-1}\hat{\G}_{k_i}.
\ee
\par Changing $u_j$ to the variables $\z_j$ defined in eq.~(\ref{zeta}), we have
\be\label{TrGaz}
\Tr[(\G_A^{T_2})^{2n}]=\left(\frac{l_0}{2}\right)^{2n}\int\displaylimits_{[0,2\pi]^{2n}}\frac{d^{2n}k}{(2\pi)^{2n}}\int\displaylimits_{R_{u}}d^{2n}\zeta D(\{k\})\tilde{F}(\{k\})e^{\mathrm{i}l_0\sum_{j=1}^{2n-1}\zeta_j(k_{j}-k_{0})/2}.
\ee
Since we are interesting in the hydrodynamic regime $l\gg 1$ and $t\gg 1$, the above integral are well approximated by the stationary phase method. The stationarity with respect to the variables $\zeta_0,\zeta_1,\cdots,\zeta_{2n-1}$ is the same with last section
\be
k_j=k_0,\qquad \forall j\in[1,2n-1].
\ee
We can replace any $k_j$ with $k_0$ except highly oscillating terms: $e^{-2\mathrm{i}\varepsilon_{k_i}t\s_y}$ in $\hat{\G}_{k_i}$ and terms like $\sin[(k_i-k_{i+1})l_0/2]$ in the first factor of $\tilde{F}(\{k\})$ in eq.~{\ref{tF}}. Again, one can ignore the term $D(\{k\})$ in the integral since $D(\{k\})=1$ at the stationary point.
\par Moreover, since the integrand in eq.~(\ref{TrGaz}) does not depend on $\zeta_0$, we can first integrate out $\zeta_0$ obtaining
\be
\Tr[(\G_A^{T_2})^{2n}]=\left(\frac{l_0}{2}\right)^{2n}\int\displaylimits_{[0,2\pi]^{2n}}\frac{d^{2n}k}{(2\pi)^{2n}}\int d^{2n-1}\zeta ~\mu(\{\zeta\})\tilde{F}(\{k\})e^{\mathrm{i}l_0\sum_{j=1}^{2n-1}\zeta_j(k_{j}-k_{0})/2}.
\ee
\par Now we can apply stationary phase approximation to evaluate the integral in the $4n-2$ variables $k_1,\cdots,k_{2n-1}$ and $\z_1,\cdots,\z_{2n-1}$.
\par Noting that
\be\label{sin}
(-1)^n\prod_{j=0}^{2n-1}2\sin\left[\frac{(k_j-k_{j+1})l_0}{2}\right]=2+\sum_{p=1}^{2n}(-1)^p\sum_{j_1<\cdots<j_{p}=0}^{2n-1}e^{\mathrm{i}l_0\sum_{i=1}^p(k_{j_i}-k_{j_{i}+1})}+e^{-\mathrm{i}l_0\sum_{i=1}^p(k_{j_i}-k_{j_{i}+1})},
\ee
we find that in this case, $\tilde{F}(\{k\})$ get contributed from two parts. The first part is attributed to the constant term at the right-hand side of eq.~(\ref{sin}), and the result is given by $2F(\{k\})$ in eq.~(\ref{F}). The second part comes from the product of the summation term at the right-hand side of eq.~(\ref{sin}) with $\Tr\prod_{i=0}^{2n-1}\hat{\G}_{k_i}$. The contribution of the first part has already been worked out in the last section (c.f. eq.~(\ref{TrGaf}) with $l$ replaced by $l_0$). In the second part of $\tilde{F}(\{k\})$, taking care of the properties of the measure $\m(\{\z\})$, one finds that only when the terms of eq.~(\ref{F1}) and eq.~(\ref{sin}) that contain the same quasimomenta $k_i$ are multiplied, one can get a non-zero $\m(\{\z\})$.
\par The additional stationary points are
\be
\begin{split}
&k_j^*=k_0,\qquad\qquad \qquad j=1,2,\cdots,2n-1\\
&\z_j^*=(-1)^j(\pm2-4\varepsilon'(k_0) t/l_0),\qquad j=1,2,\cdots,2p-1\\
&\z_j^*=0 \qquad\qquad \qquad j=2p,\cdots,2n-1.
\end{split}
\ee
Noting that the $\pm$ sign in the equation above, there are two stationary points contribute to the measure. Adding them together, the total measure at the stationary points above is given by $\m(\{\z_j^*\})=2g_2(k_0,t/l_0)$, where
\be
g_2(k,x)=2|v(k)|x+2\max(|v(k)|x,1)-2\max(2|v(k)|x,1).
\ee
The function $g_2(k,t/l_0)$ describes a linear growth up to $t=l_0/(2|v(k)|)$ and a linear decrease followed till $t=l_0/|v(k)|$. At late times, it vanishes.
\par Then we have
\be\label{TrGaT2f}
\Tr[(\G_A^{T_2})^{2n}]=2la^{2n}+2l[(a^2+b^2)^n-a^{2n}]\int_0^{2\pi}\frac{dk}{2\pi}\Big[g_1(k,t/l_0)+\frac12g_2(k,t/l_0)\Big].
\ee
The equation above can be simplified as
\be\label{TrGaT2f}
\Tr[(\G_A^{T_2})^{2n}]=2la^{2n}+2l[(a^2+b^2)^n-a^{2n}]\int_0^{2\pi}\frac{dk}{2\pi}\max(0,1-|v(k)|t/l_0),
\ee
where we have used the fact
\be
g_1(k,x)+\frac12 g_2(k,x)=g_1(k,x/2)=\max(0,1-|v(k)|x).
\ee
\par For R\'enyi negativity with index $\a$, we have
\be\label{TrGaF}
\begin{split}
\mE_{\a}=2l(1-\a)\mF_{\a}(a^2)+2l(1-\a)[\mF_{\a}(a^2+b^2)-\mF_{\a}(a^{2})]\int_0^{2\pi}\frac{dk}{2\pi}\max(0,1-|v(k)|t/l_0).
\end{split}
\ee
Alternatively, using the notation $s_{\mathrm{q}}^{(\a)}$ and $s_{\mathrm{mix}}^{(\a)}$ introduced in eq.~(\ref{sq}) and eq.~(\ref{sm}) respectively, we can write the R\'enyi negativity with index $\a$ as
\be
\mE_{\a}/(1-\a)=\int\frac{dk}{2\pi}(s^{(\a)}_{\mathrm{q}}-s^{(\a)}_{\mathrm{mix}})\min(l,2|v(k)|t)+l\int\frac{dk}{2\pi}s^{(\a)}_{\mathrm{mix}}.
\ee
\par From the expression above we know that $\mE_{\a}$ are not good measures of entanglement between $A_1$ and $A_2$, since the function $\min(l,2|v(k)|t)$ does not count the number of entangled pairs of quasi-particles between $A_1$ and $A_2$. Instead, it represent the number of pairs of entangled quasi-particles with quasi-momenta $k$
and $-k$ that are shared between the full subsystem $A=A_1\cup A_1$ and the rest at time $t$.
\par Carrying out the integral above, we obtain
\be\label{Ea}
\mE_{\a}=2l(1-\a)\mF_{\a}(a^2)\mG(|g|t/l)+2l(1-\a)\mF_{\a}(a^2+b^2)[1-\mG(|g|t/l)].
\ee
\begin{figure}
        \centering
        \subfloat
        {\includegraphics[width=7cm]{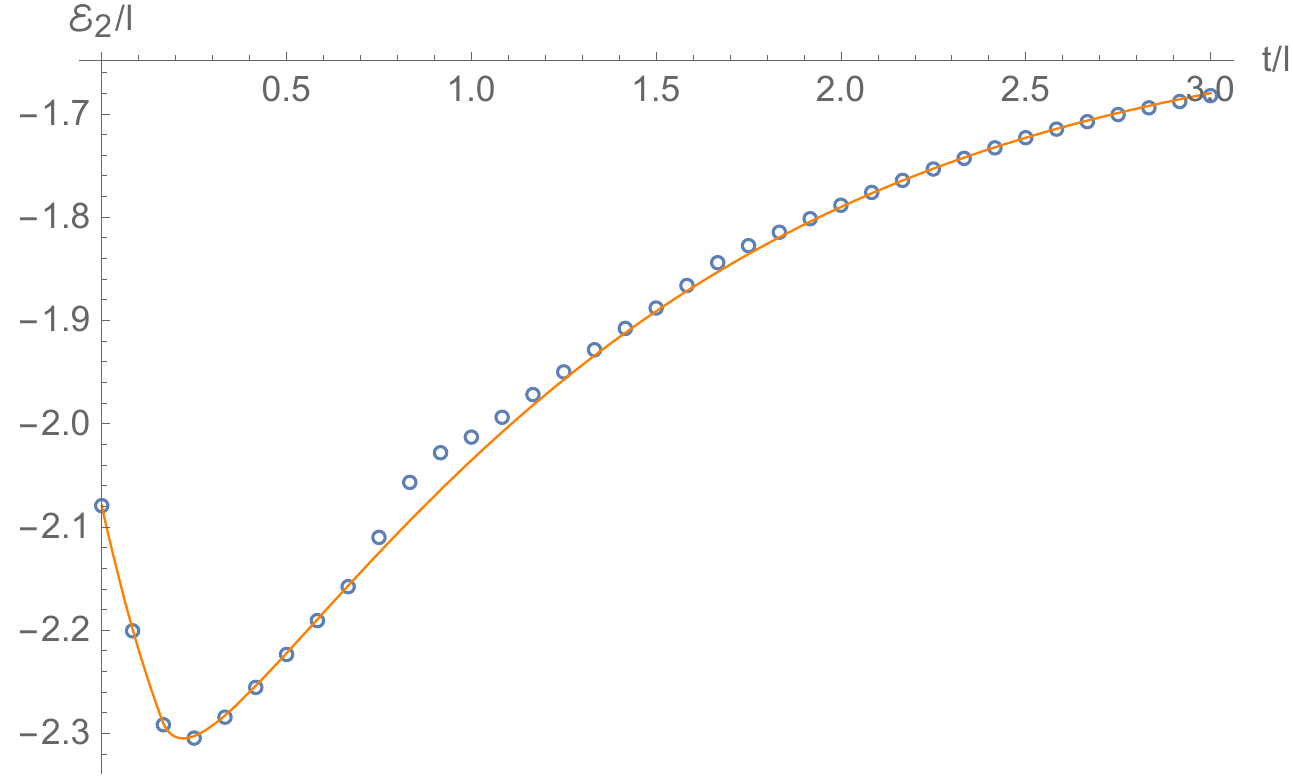}} \quad\quad
        {\includegraphics[width=7cm]{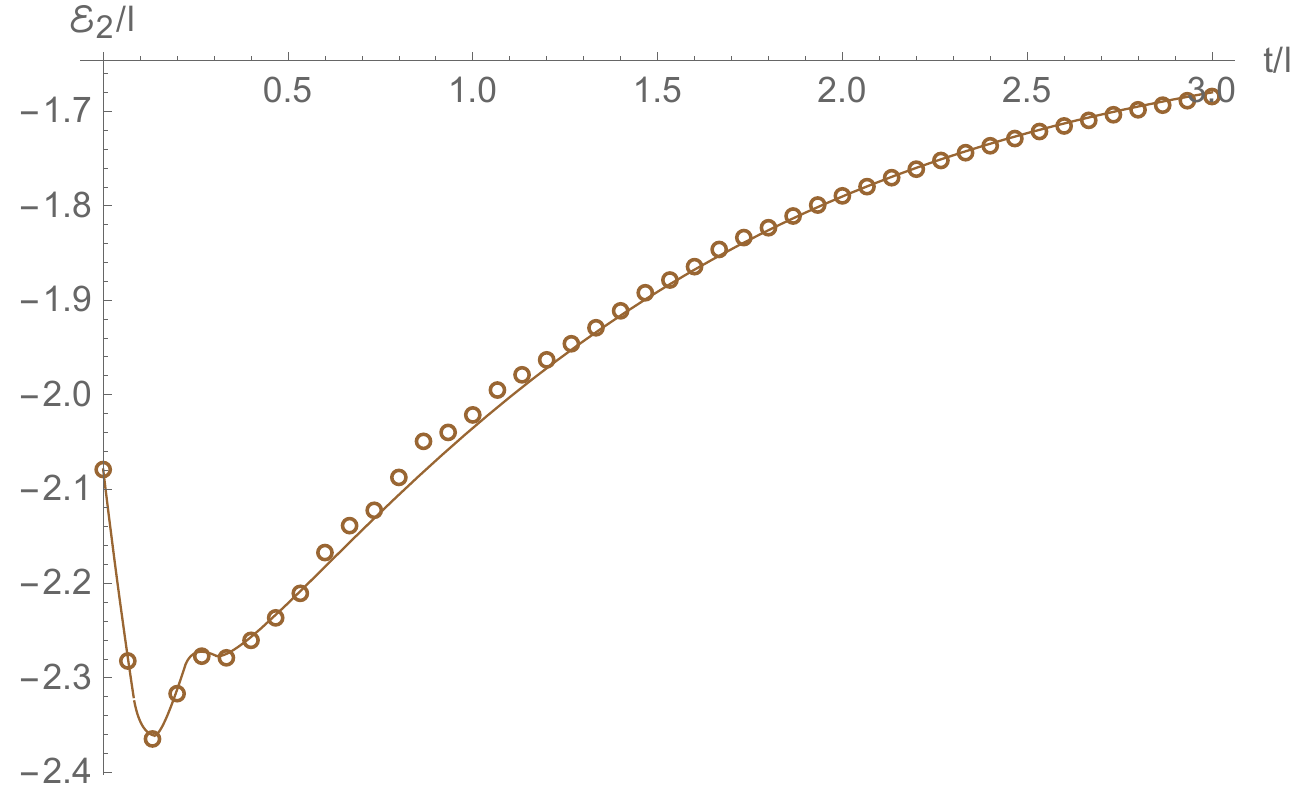}}
        \caption{R\'enyi negativity $\mathcal{E}_2$ as a function of $t/l$. The full lines are the quasi-particle predictions (cf. eq.~(\ref{Ea}) and eq.~(\ref{Ed})).Left panel:adjacent intervals. Right panel: disjoint intervals with $d=50$. The parameters we choose are $L=300,l_0=30$, $\g^+=l_0^{-1},\g^-=1.5l_0^{-1},\e=0.2,g=3,r=0.5,\eta=4$. As shown in the figure, numerical data matches the analytical result very well.}
        \label{fig2}
\end{figure}
\subsection{Disjoint interval}
It's easy to generalize the results obtained above to the case where the two regions of our subsystem are not adjacent. If we denote the distance between $A_1$ and $A_2$ as $d$, then keeping track of this tiny difference in the derivation in the last subsection, one concludes that in this case, the stationary points are
\be
\begin{split}
&k_j^*=k_0,\qquad\qquad \qquad j=1,2,\cdots,2n-1\\
&\z_j^*=(-1)^j\left(\pm2\frac{l_0+d}{l_0}-4\varepsilon'(k_0)\frac{t}{l_0}\right),\qquad j=1,2,\cdots,2p-1\\
&\z_j^*=0 \qquad\qquad \qquad j=2p,\cdots,2n-1.
\end{split}
\ee
the correspond measure $\m(\vec{\z}^*)$ now becomes $2\tilde{g}_2(k_0,t/l_0)$, where
\be
\tilde{g}_2(k,x)=\max(2|v(k)|x,2+d/l_0)+\max(2|v(k)|x,d/l_0)-2\max(2|v(k)|x,1+d/l_0).
\ee
Then in this case, the R\'enyi negativity with index $\a$ is given by
\be\label{TrGaF1}
\begin{split}
\mE_{\a}=2l(1-\a)\mF_{\a}(a^2)+2l(1-\a)[\mF_{\a}(a^2+b^2)-\mF_{\a}(a^{2})]\int_0^{2\pi}\frac{dk}{2\pi}\Big[g_1(k,t/l_0)+\frac12\tilde{g}_2(k,t/l_0)\Big].
\end{split}
\ee
Similarly, we can express $\mE_{\a}$ as
\be
\begin{split}
&\mE_{\a}/(1-\a)=\int\frac{dk}{2\pi}\Big[(s^{(\a)}_{\mathrm{q}}-s^{(\a)}_{\mathrm{mix}})\big(\min(l,4|v(k)|t)-\max(2|v(k)|t,l+d)\\
&-\max(2|v(k)|t,d)+2\max(2|v(k)|t,l_0+d)\big)\Big]+l\int\frac{dk}{2\pi}s^{(\a)}_{\mathrm{mix}}.
\end{split}
\ee
The integration can be worked out in terms of the $\mG$ function defined in eq.~(\ref{mG}). Hence for the reader's convenience, we report the final result here though they are cumbersome and not very illustrating. The final result of the R\'enyi negativity with index $\a$ is
\be\label{Ed}
\begin{split}
&\mE_{\a}=2l(1-\a)\mF_{\a}(a^2)+2l(1-\a)[\mF_{\a}(a^2+b^2)-\mF_{\a}(a^{2})]\\
&\times\Big[1-\mG\big(\frac{|g|t}{l_0}\big)+(1+\frac{d}{l_0})\mG\big(\frac{|g|t}{l_0+d}\big)-\big(1+\frac{d}{2l_0}\big)\mG\big(\frac{|g|t}{2l_0+d}\big)-\frac{d}{2l_0}\mG\big(\frac{|g|t}{d}\big)\Big].
\end{split}
\ee
\par In our case, since $\sqrt{a^2}>1/2$ and $\sqrt{a^2+b^2}>1/2$ for all $t$, the entanglement negativity vanishes identically.
\begin{figure}
        \centering
        \subfloat
        {\includegraphics[width=7cm]{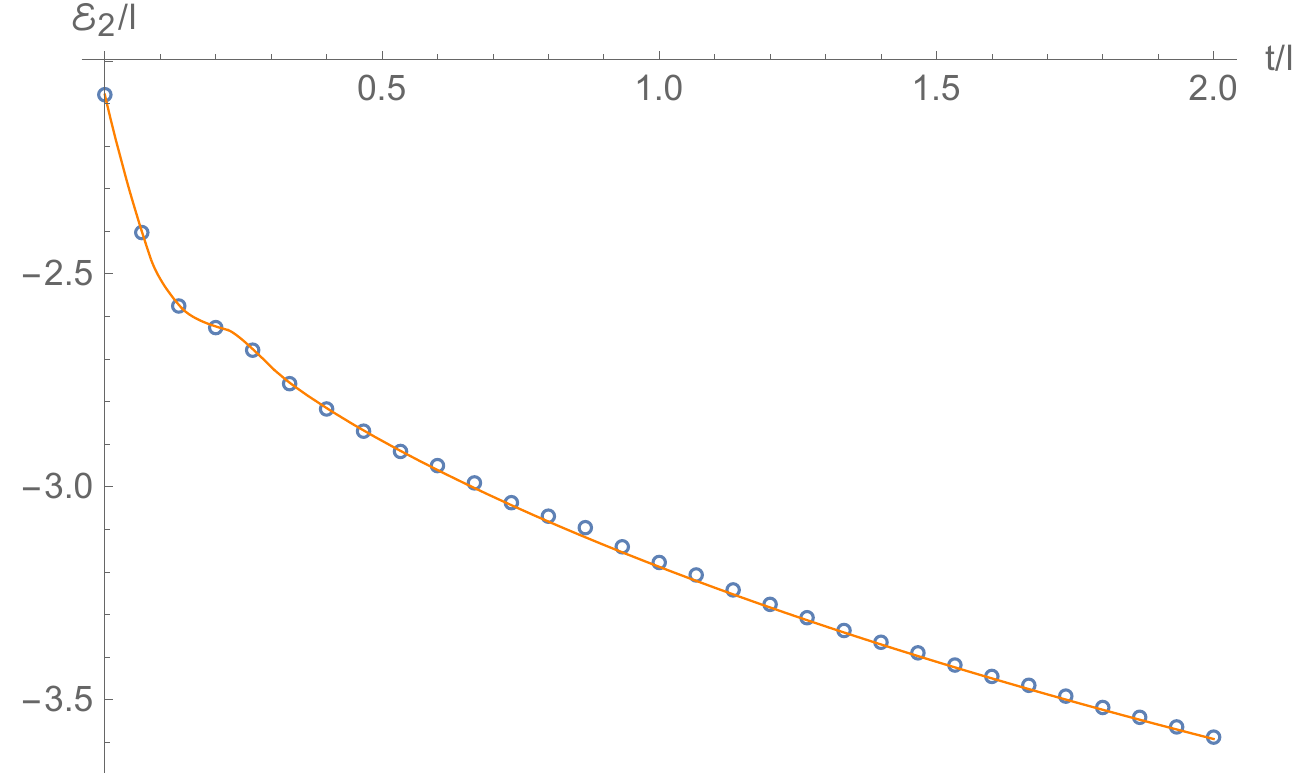}} \quad\quad
        {\includegraphics[width=7cm]{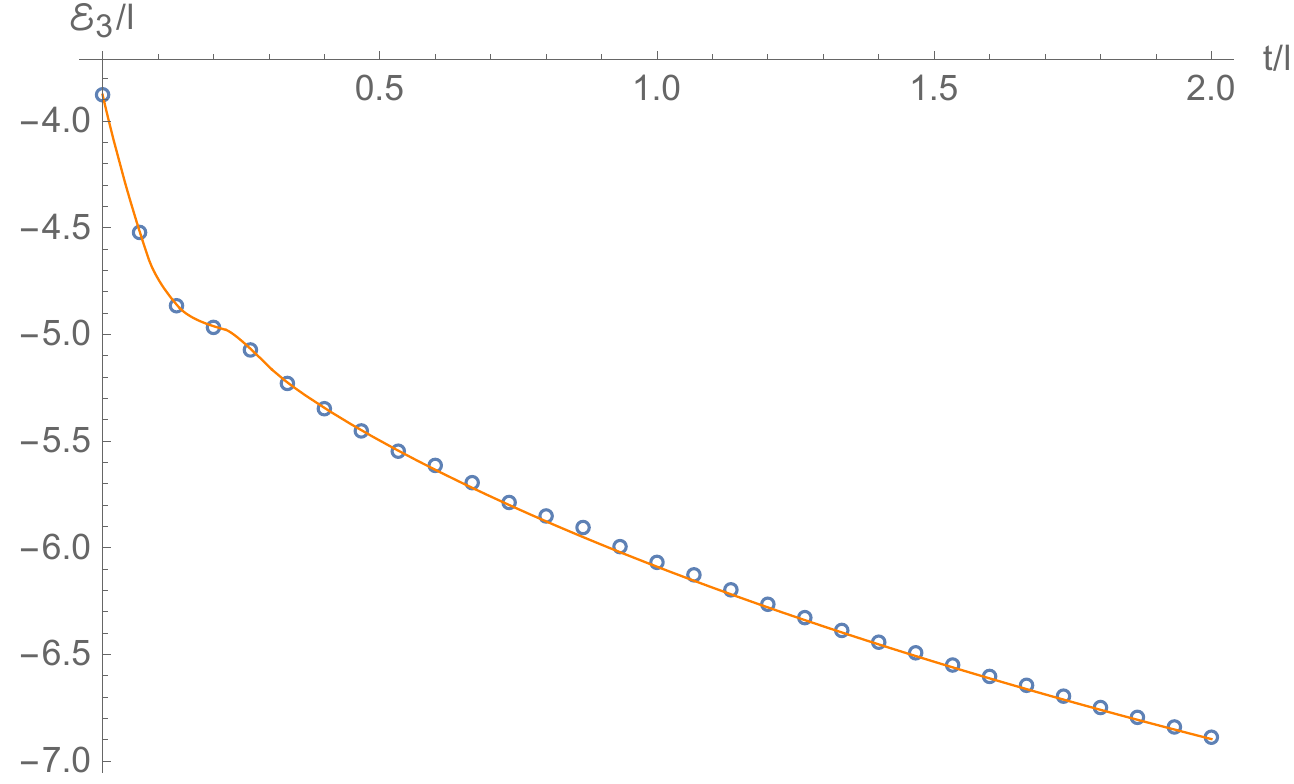}}
        \caption{R\'enyi negativity $\mathcal{E}_{\a}$ as a function of $t/l$ for disjoint intervals. The full lines are the quasi-particle predictions (cf. eq.~(\ref{Ea}) and eq.~(\ref{Ed})). Left panel: $\mathcal{E}_{2}$. Right panel: $\mathcal{E}_{3}$. Here we consider the critical case $\g^+=\g^-=l_0^{-1}$ where there is no steady state at late times. The other parameters we choose are $L=300,l_0=30,d=50$, $\e=0.2,g=3,r=0.5,\eta=4$. As shown in the figure, the agreement is extremely excellent.}
        \label{fig3}
\end{figure}
\par As shown in fig.~(\ref{fig2}), even for disjoint interval case, $\mE_{\a}$ exhibits a linear behavior at early times. But we know that there are no shared entangled pairs between $A_1$ and $A_2$ when $t\ll d$. This reflect the fact that $\mE_{\a}$ are not proper measures of entanglement between $A_1$ and $A_2$. 
\section{Numerical benchmark}
The $2L\times 2L$ matrix $C$ (cf. eq.~(\ref{C}))is complex, we write it as $C=\mR+\mathrm{i}\mI$. Define $U(t)=e^{t\S(h+\mI)}$, the evolution of the covariance matrix is \cite{carollo2022dissipative}
\be
G(t)=U(t)G(0)U(t)^{T}+\int_0^{t}dt'U(t')\S\mR\S^{T}U(t')^{T}.
\ee
\par If we denote the eigenvalues of the matrix $\mathrm{i}\S_A G_A(t)$ as $\{\pm\l_1(t),\pm\l_2(t),...,\pm\l_l(t)\}$, then the moments of the RDM are obtained by
\be
\Tr[(\rho_A(t))^{\a}]=\prod_{i=1}^{l}\left[\left(\l_i(t)+\frac12\right)^{\a}-\left(\l_i(t)-\frac12\right)^{\a}\right]^{-1}.
\ee
From the equation above, the R\'enyi entropies are obtained straightforwardly.
\par In bosonic systems, the net effect of partial transposition with respect to $A_2$ is changing the sign of the momenta corresponding to $A_2$. Thus the momenta correlators in the partially transposed density matrix can be obtained by simply changing the sign of the momenta in $A_2$ (cf. eq.~(\ref{GA})).
If we denote the eigenvalues of $\mathrm{i}\S_AG_A(t)^{T_2}$ by $\{\tau_1^2(t),\tau_2^2(t),\cdots,\tau_l^2(t)\}$, then the  R\'enyi negativity with index $\a$ is given by
\be
\mathcal{E}_{\a}(t)=-\sum_{j=1}^l\log\Big[\lt(\t_j(t)+\frac12\rt)^{\a}-\lt(\t_j(t)-\frac12\rt)^{\a}\Big].
\ee
\par The numerical data of the dynamics of the R\'enyi negativities are shown in fig.~\ref{fig1}, fig.~\ref{fig2} and fig.~\ref{fig3}, in which the analytical predictions are drawn with the full line for comparison.
\section{Conclusion}\label{section8}
In this paper, we discuss the temporal evolution of R\'enyi negativities after global quenches in free-boson chain with homogeneous local linear dissipation. Firstly, we consider a simple example where the initial state is a thermal state. The time dependence of R\'enyi entropies can be obtained directly. When we initially prepared the system in the squeezed thermal state, the computation becomes more involved, and more sophisticated approaches are needed. Since we are interested in the hydrodynamic regime, we adopt the stationary phase approximation to evaluate the dynamics of R\'enyi negativity. Finally, we obtained the exact formula and test our results against numerical computation and found they match very well.
\par It would be very interesting to investigate the evolution of entanglement negativity in dissipative interacting integrable models. It would be also interesting to see the interplay between symmetry and dissipation, i.e. to study the symmetry resolved entanglement in free boson chains with dissipation when the initial state is symmetric \cite{murciano2023symmetry}. When the initial state breaks the symmetry, then one should consider the dynamics of entanglement asymmetry instead to see whether the broken symmetry can be restored at late times along the line in \cite{ares2023entanglement, ares2023lack}.
\begin{appendix}
\section{Squeezed thermal state}
In this section, we will briefly review some basic properties of the squeezed thermal state. The squeezing operator defined in eq.~(\ref{squ}) is unitary $\mS^{\dg}\mS=1$. For the single-mode squeezing operator $\mS_1$, one has
\be
\begin{split}
\mS_1(r,\phi)a\mS_1^{\dg}(r,\phi)=a\cosh r+e^{2\mathrm{i}\phi}a^{\dg}\sinh r,\\
\mS_1(r,\phi)a^{\dg}\mS_1^{\dg}(r,\phi)=a^{\dg}\cosh r+e^{-2\mathrm{i}\phi}a\sinh r.
\end{split}
\ee
\par For the thermal state
\be
\rho_{th}=\frac{1}{Z}e^{-\beta H_0}=\prod_{i=1}^{L}\frac{1}{1-e^{-\beta\e}}e^{-\beta\e\hat{n}_i}.
\ee
We have
\be
\begin{split}
\Tr(\rho_{th}a_i^{\dg}a_j)=\d_{ij}\frac{1}{1-e^{-\beta\e}}\sum_{n=0}^{\infty}ne^{-n\beta\e}\\
=-\d_{ij}\frac{\partial}{\partial\beta}\log\frac{1}{1-e^{-\beta\e}}=\d_{ij}\frac{1}{e^{\beta\e}-1}=\d_{ij}\bar n.
\end{split}
\ee
Then using the canonical commutation relation $[a_i,a_j^{\dg}]=\d_{ij}$, we obtain
\be
\Tr(\rho_{th}a_ia_j^{\dg})=\d_{ij}(1+\bar n).
\ee
All the operators which have non-vanishing expectation values in thermal state must be functions of $\hat{n}_i$.
\par For a initial squeezed thermal state, we have
\be
\begin{split}
&\Tr[\rho_0a^{\dg}_ia^{\dg}_j]=\Tr[\mS\rho_{th}\mS^{\dg}a^{\dg}_ia^{\dg}_j]=\Tr[\rho_{th}\mS_1a^{\dg}_i\mS_1^{\dg}\mS_1a^{\dg}_j\mS_1^{\dg}]\\
&=e^{-2\mathrm{i}\phi}\sinh r\cosh r[\Tr(\rho_{th}a_i^{\dg}a_j)+\Tr(\rho_{th}a_ia_j^{\dg})]\\
&=\d_{ij}(2\bar n+1)e^{-2\mathrm{i}\phi}\sinh r\cosh r.
\end{split}
\ee
In a similar way, one can obtain
\be
\begin{split}
&\Tr[\rho_0a^{\dg}_ia_j]=\d_{ij}[\bar n\cosh^2r+(\bar n+1)\sinh^2r],\\
&\Tr[\rho_0a_ia^{\dg}_j]=\d_{ij}[(\bar n+1)\cosh^2r+\bar n\sinh^2r],\\
&\Tr[\rho_0a_ia_j]=\d_{ij}(2\bar n+1)e^{2\mathrm{i}\phi}\sinh r\cosh r.
\end{split}
\ee
\par Alternatively, one can use the characteristic function of the squeezed thermal state
\be
\chi(\bm{\a})\equiv\Tr(\rho_0D_{\bm{\a}}),
\ee
where $D_{\bm{\a}}$ is the displacement operator
\be
D_{\bm{\a}}=\prod_ie^{\a_i a_i-\a_i^*a_i^{\dg}}.
\ee
For the squeezed thermal state, the characteristic function is given by
\be
\chi(\bm{\a})=\prod_{i}\exp\left[-\frac12\coth\left(\frac{\beta\e}{2}\right)\big|\a_i\cosh r+\a_i^*\sinh r e^{2\mathrm{i}\phi}\big|^2\right].
\ee
Then the expectation value of a product of arbitrary numbers of $a$ and $a^{\dg}$ can be computed as
\be
\Tr(\rho_0a_i^{\dg m}a_j^n)=\left(\frac{\p}{\p\a_i}\right)^n\left(-\frac{\p}{\p\a_j^*}\right)^m\chi(\bm{\a})\Big|_{\bm{\a}=\bm{\a}^*=0}.
\ee
\par The occupation number in momentum space is obtained from the position space density by Fourier transform
\be\label{n0}
\Tr(\rho_0b_k^{\dg}b_k)=\frac{1}{L}\sum_{i,j=1}^{L}e^{\mathrm{i}(i-j)k}\Tr(\rho_0a^{\dg}_ia_j)=\bar n \cosh(2r)+\sinh^2r.
\ee
\end{appendix}
\section*{Acknowledgments}
This work was supported  by the National Natural Science Foundation of China, Grant No.\ 12005081.
\bibliography{2022}

\begin{thebibliography}{10}

\bibitem{Amico:2007ag}
L.~Amico, R.~Fazio, A.~Osterloh, and V.~Vedral, ``{Entanglement in many-body
  systems},'' {\em Rev. Mod. Phys.}, vol.~80, pp.~517--576, 2008.

\bibitem{Calabrese:2009qy}
P.~Calabrese and J.~Cardy, ``{Entanglement entropy and conformal field
  theory},'' {\em J. Phys. A}, vol.~42, p.~504005, 2009.

\bibitem{Eisert:2008ur}
J.~Eisert, M.~Cramer, and M.~B. Plenio, ``{Area laws for the entanglement
  entropy - a review},'' {\em Rev. Mod. Phys.}, vol.~82, pp.~277--306, 2010.

\bibitem{Nishioka:2009un}
T.~Nishioka, S.~Ryu, and T.~Takayanagi, ``{Holographic Entanglement Entropy: An
  Overview},'' {\em J. Phys. A}, vol.~42, p.~504008, 2009.

\bibitem{Ryu:2006bv}
S.~Ryu and T.~Takayanagi, ``{Holographic derivation of entanglement entropy
  from AdS/CFT},'' {\em Phys. Rev. Lett.}, vol.~96, p.~181602, 2006.

\bibitem{Hawking:1974sw}
S.~W. Hawking, ``{Particle Creation by Black Holes},'' {\em Commun. Math.
  Phys.}, vol.~43, pp.~199--220, 1975.
\newblock [Erratum: Commun.Math.Phys. 46, 206 (1976)].

\bibitem{Hawking:1976ra}
S.~W. Hawking, ``{Breakdown of Predictability in Gravitational Collapse},''
  {\em Phys. Rev. D}, vol.~14, pp.~2460--2473, 1976.

\bibitem{Almheiri:2020cfm}
A.~Almheiri, T.~Hartman, J.~Maldacena, E.~Shaghoulian, and A.~Tajdini, ``{The
  entropy of Hawking radiation},'' 6 2020.

\bibitem{Calabrese:2004eu}
P.~Calabrese and J.~L. Cardy, ``{Entanglement entropy and quantum field
  theory},'' {\em J. Stat. Mech.}, vol.~0406, p.~P06002, 2004.

\bibitem{Peres:1996dw}
A.~Peres, ``{Separability criterion for density matrices},'' {\em Phys. Rev.
  Lett.}, vol.~77, pp.~1413--1415, 1996.

\bibitem{vidal2002computable}
G.~Vidal and R.~F. Werner, ``Computable measure of entanglement,'' {\em
  Physical Review A}, vol.~65, no.~3, p.~032314, 2002.

\bibitem{plenio2005logarithmic}
M.~B. Plenio, ``Logarithmic negativity: a full entanglement monotone that is
  not convex,'' {\em Physical review letters}, vol.~95, no.~9, p.~090503, 2005.

\bibitem{alba2021spreading}
V.~Alba and F.~Carollo, ``Spreading of correlations in markovian open quantum
  systems,'' {\em Physical Review B}, vol.~103, no.~2, p.~L020302, 2021.

\bibitem{carollo2022dissipative}
F.~Carollo and V.~Alba, ``Dissipative quasiparticle picture for quadratic
  markovian open quantum systems,'' {\em Physical Review B}, vol.~105, no.~14,
  p.~144305, 2022.

\bibitem{alba2022logarithmic}
V.~Alba and F.~Carollo, ``Logarithmic negativity in out-of-equilibrium open
  free-fermion chains: an exactly solvable case,'' {\em arXiv preprint
  arXiv:2205.02139}, 2022.

\bibitem{d2022logarithmic}
A.~D'Abbruzzo, V.~Alba, and D.~Rossini, ``Logarithmic entanglement scaling in
  dissipative free-fermion systems,'' {\em Physical Review B}, vol.~106,
  no.~23, p.~235149, 2022.

\bibitem{Calabrese:2005in}
P.~Calabrese and J.~L. Cardy, ``{Evolution of entanglement entropy in
  one-dimensional systems},'' {\em J. Stat. Mech.}, vol.~0504, p.~P04010, 2005.

\bibitem{vidmar2016generalized}
L.~Vidmar and M.~Rigol, ``Generalized gibbs ensemble in integrable lattice
  models,'' {\em Journal of Statistical Mechanics: Theory and Experiment},
  vol.~2016, no.~6, p.~064007, 2016.

\bibitem{lindblad1976generators}
G.~Lindblad, ``On the generators of quantum dynamical semigroups,'' {\em
  Communications in Mathematical Physics}, vol.~48, pp.~119--130, 1976.

\bibitem{breuer2002theory}
H.-P. Breuer and F.~Petruccione, {\em The theory of open quantum systems}.
\newblock Oxford University Press, USA, 2002.

\bibitem{alba2022hydrodynamics}
V.~Alba and F.~Carollo, ``Hydrodynamics of quantum entropies in ising chains
  with linear dissipation,'' {\em Journal of Physics A: Mathematical and
  Theoretical}, vol.~55, no.~7, p.~074002, 2022.

\bibitem{calzetta2009nonequilibrium}
E.~A. Calzetta and B.-L.~B. Hu, {\em Nonequilibrium quantum field theory}.
\newblock Cambridge University Press, 2009.

\bibitem{calabrese2012quantum}
P.~Calabrese, F.~H. Essler, and M.~Fagotti, ``Quantum quench in the transverse
  field ising chain: I. time evolution of order parameter correlators,'' {\em
  Journal of Statistical Mechanics: Theory and Experiment}, vol.~2012, no.~07,
  p.~P07016, 2012.

\bibitem{murciano2023symmetry}
S.~Murciano, P.~Calabrese, and V.~Alba, ``Symmetry-resolved entanglement in
  fermionic systems with dissipation,'' {\em arXiv preprint arXiv:2303.12120},
  2023.

\bibitem{ares2023entanglement}
F.~Ares, S.~Murciano, and P.~Calabrese, ``Entanglement asymmetry as a probe of
  symmetry breaking,'' {\em Nature Communications}, vol.~14, no.~1, p.~2036,
  2023.

\bibitem{ares2023lack}
F.~Ares, S.~Murciano, E.~Vernier, and P.~Calabrese, ``Lack of symmetry
  restoration after a quantum quench: an entanglement asymmetry study,'' {\em
  arXiv preprint arXiv:2302.03330}, 2023.

\end{thebibliography}
\bibliographystyle{ieeetr}
\end{document}